\documentclass[%
 reprint,
 amsmath,amssymb,
 aps,
]{revtex4-1}
\usepackage{float}
\usepackage[normalem]{ulem}
\usepackage{graphicx}
\usepackage{dcolumn}
\usepackage{bm}
\usepackage[dvipsnames]{xcolor}
\usepackage{filecontents}

\newcommand{\tr}{\operatorname{tr}}

\newcommand{\ket}[1]{\vert #1 \rangle}
\newcommand{\braket}[1]{\langle #1 \rangle}

\renewcommand{\vec}[1]{{\mathbf #1}}

\newcommand{\TJWat}{IBM Quantum, IBM T.J. Watson Research Center, Yorktown Heights, NY 10598, USA}

\begin{document}


\title{Scalable error mitigation for noisy quantum circuits produces competitive expectation values}

\author{Youngseok Kim}
\author{Christopher J. Wood}
\author{Theodore J. Yoder}
\author{Seth T. Merkel}
\author{Jay M. Gambetta}
\author{Kristan Temme}
\author{Abhinav Kandala}

\affiliation{\TJWat}
\date{\today}

\pacs{Valid PACS appear here}
\maketitle


{\bf{Noise in existing quantum processors only enables an approximation to ideal quantum computation. However, these approximations can be vastly improved by error mitigation~\cite{Kristan2017,Li2017}, for the computation of expectation values, as shown by small-scale experimental demonstrations~\cite{kandala2019error,dumitrescu2018,song2019,zhang2020errormit}. However, the practical scaling of these methods to larger system sizes remains unknown. Here, we demonstrate the utility of zero-noise extrapolation~\cite{Kristan2017,Li2017,kandala2019error} for relevant quantum circuits using up to 26 qubits, circuit depths of 60, and 1080 CNOT gates. We study the scaling of the method for canonical examples of product states and entangling Clifford circuits of increasing size, and extend it to the quench dynamics of 2-D Ising spin lattices with varying couplings. We show that the efficacy of the error mitigation is greatly enhanced by additional error suppression techniques and native gate decomposition that reduce the circuit time. By combining these methods, we demonstrate an accuracy in the approximate quantum simulation of the quench dynamics that surpasses the classical approximations obtained from a state of the art 2-D tensor network method. These results reveal a path to a relevant quantum advantage with noisy, digital, quantum processors.}}

Decoherence and unitary gate errors currently limit the volume and fidelity of quantum circuits. While this can be remedied with the advent of quantum error correction~\cite{shor1995,Steane1996}, a fully fault-tolerant hardware architecture is not immediately accessible. Although existing quantum processors~\cite{zhang2020,arute2019,wu2021strong} have achieved a size that pushes at the boundaries of classical simulability, it is important to ask if such noisy machines can perform useful computations. In this context, recent theoretical work has proven that even noisy shallow depth quantum circuits can outperform their noiseless classical counterparts~\cite{bravyi2020noisy}. Furthermore, recently proposed error mitigation techniques~\cite{Kristan2017,Li2017,kandala2019error} present a path to obtain accurate expectation values even on noisy quantum computers. The general operating principle of these techniques is to reconstruct a noise-free estimate of an expectation value, from multiple noisy experiments. These techniques are particularly attractive for the near-term because they typically require no additional overhead for the most valuable resources - qubit number and circuit depth. While error mitigation only improves the estimation of expectation values, such computations represent a sizable portion of near-term quantum algorithms, for instance, the estimation of molecular energies ~\cite{Peruzzo2014,kandala2017hardware} in variational quantum eigensolvers, or kernel estimation in machine learning~\cite{havlivcek2019supervised,Schuld2019}. This has motivated a number of theoretical proposals and experimental demonstrations of error mitigation, that can be broadly categorized into general-purpose~\cite{Kristan2017,Li2017,kandala2019error,dumitrescu2018,song2019,zhang2020errormit,koczor2021exponential,huggins2021virtual} and problem-specific techniques~\cite{McClean_2017,Bonet_Monroig_2018,McArdle_2019}.

A widely adopted general-purpose error mitigation technique is zero-noise extrapolation. Here, expectation values are measured for circuits run at varying noise levels to extrapolate to its zero noise-limit. The noise levels in the circuit are typically varied either by stretching the control pulses in time~\cite{Kristan2017,kandala2019error}, or by the insertion of noisy identity-equivalent operations~\cite{dumitrescu2018,He2020}. Recent extensions also involve learning the noise amplification factors with near-Clifford training circuits~\cite{lowe2020,czarnik2021}. Initial small-scale experiments with just four qubits of a fixed-frequency superconducting transmon~\cite{koch2007} processor and stretching of pulses in entangling cross-resonance gates~\cite{chow2011} demonstrated the promise of the technique - enabling variational ansatz with increased circuit depth to achieve ground-state energies with far-improved accuracy for molecular and interacting spin Hamiltonians~\cite{kandala2019error}. However, the hardware-platform has witnessed tremendous improvements since this initial demonstration. The size of the largest quantum processors based on this architecture now lies at 65 qubits~\cite{zhang2020}, reported transmon coherence times are exceeding several hundred microseconds~\cite{gordon2021}, cross-resonance gate fidelities are approaching 99.9 $\%$ with sub-100 ns gate times~\cite{wei2021quantum}, with steady progress in holistic metrics such as quantum volume~\cite{Jurcevic_2021} as well. This raises a tantalizing question of how much these hardware improvements could extend the reach of zero-noise extrapolation to perform accurate quantum computation at a larger scale, into the realm of quantum advantage. In this context, once we approach a scale where the general performance of an error mitigation technique cannot be compared to exact numerics, it is particularly important to understand the error bounds, as provided for the methods discussed in~\cite{Kristan2017}. For zero-noise extrapolation, a loose upper bound on the error of the zero-noise estimate after $n$-th order Richardson extrapolation is $\sim\mathcal{O}((NT\lambda)^{n+1})$, where $N$ is the number of qubits, $\lambda$ is a parameter that describes the strength of the noise  (eg. $\lambda \sim T_1^{-1}$ for purely amplitude damping noise with energy relaxation time $T_1$) and $T$ is the total evolution time. Therefore, the performance of zero-noise extrapolation is expected to be limited by $NT\lambda$ being a small number. 

\begin{figure*}[!ht]
\centerline{\includegraphics[width=2\columnwidth]{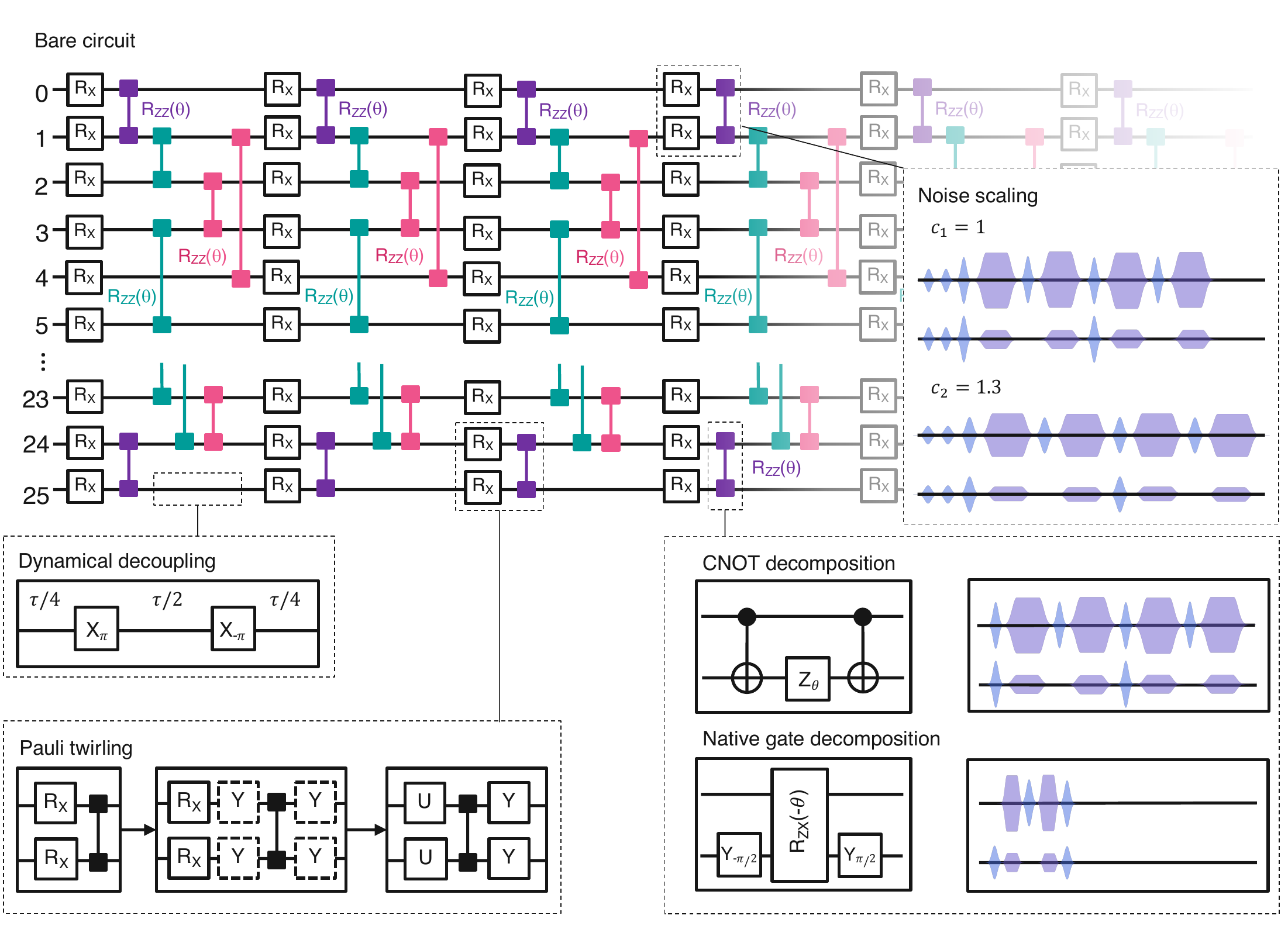}}
\caption{ {\bf {Zero-noise extrapolation for intermediate-scale short-depth quantum circuits}} An example of a 26-qubit short-depth quantum circuit composed of layers of single qubit ($R_X$) and two qubit ($R_{ZZ}$) gates that we employ to study the performance of error mitigation. The noise is amplified for zero-noise extrapolation by a appropriately scaling the duration and amplitude of the microwave pulses that compose the gates. A number of additional techniques are employed to enhance the performance of zero-noise extrapolation, that are not incorporated for the \emph{bare circuit}. The insertion of dynamical decoupling $X_{\pi}-X_{-\pi}$ sequences is used to suppress dephasing during qubit idle times. Pauli twirling averages out off-diagonal coherent errors in the Pauli basis and is implemented by sandwiching the two-qubit gates with additional single qubit gates drawn from a set of Pauli operations. 
The twirling gates can be combined with original single qubit gates into arbitrary single-qubit rotations $U(\theta,\phi,\varphi)=R_Z(\phi)R_X(-\pi/2)R_Z(\theta)R_X(\pi/2)R_Z(\varphi)$ to minimize any additional circuit time. The total evolution time is dominated by two-qubit gates, and can be reduced by native-gate decomposition. The standard decomposition of $R_{ZZ}(\theta)$ operations for arbitrary $\theta$ uses two CNOT gates, that are each constructed on our hardware using cross-resonance pulses to tune fully-entangling $R_{ZX}(\pi/2)$ operations. Alternately, the circuit time can be significantly reduced by constructing $R_{ZZ}(\theta)$ gates from partially-entangling cross-resonance pulses that implement $R_{ZX}(\theta)$.}
\label{fig:circuit}
\end{figure*}

In this work, we experimentally study the performance of zero-noise extrapolation for circuits employing up to 26 transmon qubits of a fixed-frequency quantum processor, for a maximum circuit depth of 60 with a maximum CNOT count of 1080 gates (Fig.~\ref{fig:circuit}). We initially present a couple of canonical examples - the $T_1$ decay of multi-qubit product states, and 21-qubit Greenberger–Horne–Zeilinger (GHZ) states. These experiments suggest that the realistic error bounds are far better than the loose upper bounds discussed above. Even for a fixed number of qubits $N$, the errors in the extrapolation display circuit-specific dependencies on the weight and locality of the mitigated observables. We then extend our experiments to non-Clifford circuits, studying the time dynamics of a 26 spin Ising Hamiltonian for varying system parameters. We show that the performance of zero-noise extrapolation can be significantly enhanced by suppressing $\lambda$ with additional error suppression strategies and reducing the circuit evolution time $T$ with more efficient gate decomposition and compiling, depicted in Fig.~\ref{fig:circuit}. With these improvements, we show that we can already begin to achieve an accuracy on error mitigated observables, measured off a noisy quantum processor, that is able to surpass the accuracy of an established tensor network method in the simulation of dynamics \cite{murg2007variational}. 

We perform our experiments on a superconducting processor with 27 transmon qubits, and fixed connectivity in the heavy-hex ~\cite{chamberland2020} geometry. The connectivity is favourable towards the frequency imprecision~\cite{hertzberg2020laser} and crosstalk of fixed-frequency superconducting architectures, while enabling the implementation of quantum error correcting codes~\cite{chamberland2020}. All the control and readout of the qubits is performed with microwave pulses, with entangling operations based off the cross-resonance~\cite{chow2011} effect, with a native $ZX$ interaction. Additional details of the device are provided in the Supplementary Information.

\begin{figure}[!t]
\centerline{\includegraphics[width=1\columnwidth]{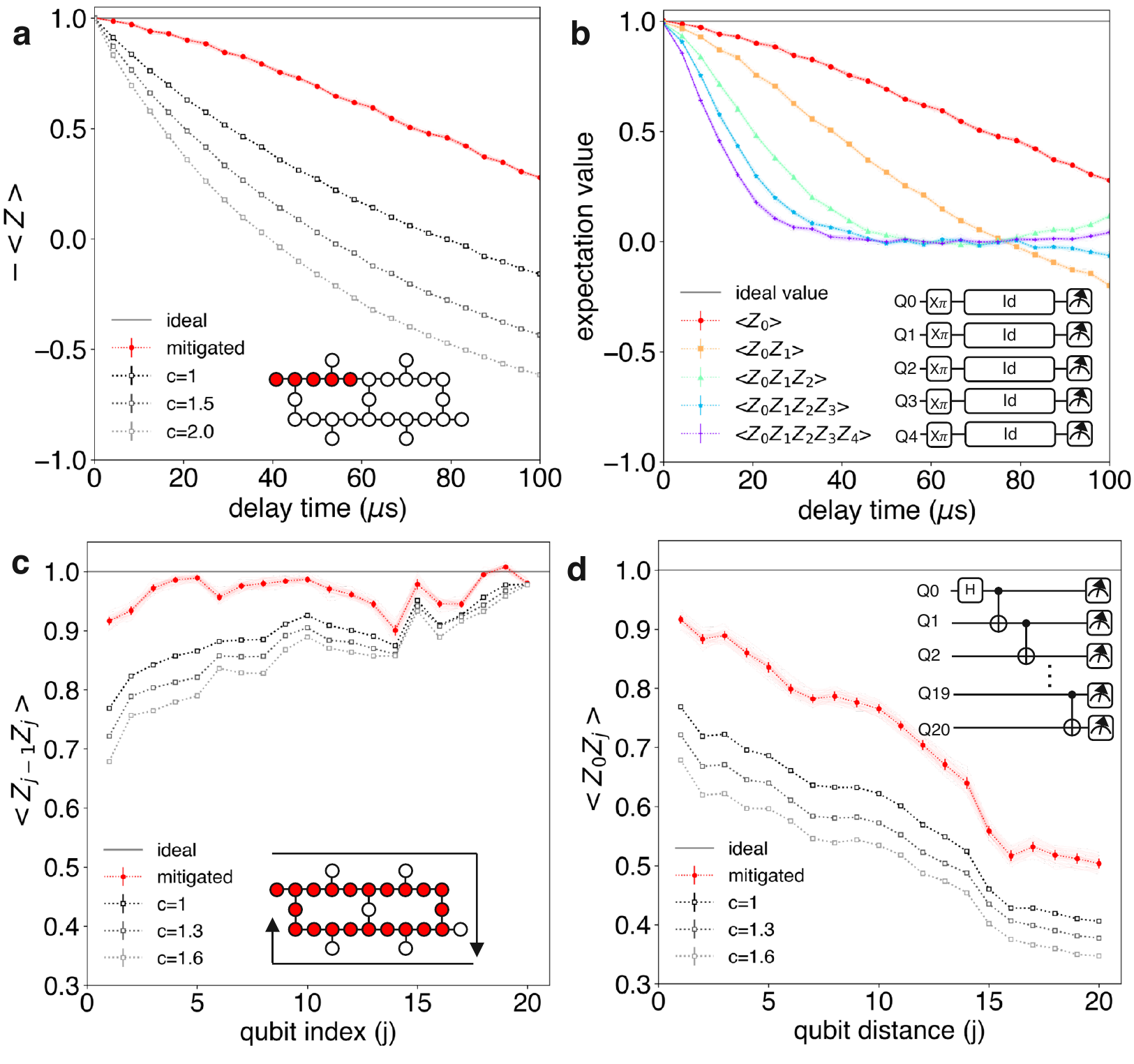}}
\caption{ {\bf {Scaling of zero-noise extrapolation for product states and entangling Clifford circuits}}
{\bf {a}} Error-mitigated (red) and unmitigated decay of $\braket{Z}$ for a standard $T_1$ decay circuit. The unmitigated experiments are performed for three different stretch factors, $c_i=1,1.5,2$, by stretching the $T_1$ delay times (denoted as Id in inset of {\bf b}). The mitigated curve is obtained by a linear extrapolation of the observables measured from the stretched experiments. Inset depicts the device connectivity and highlights the five qubits employed for the $T_1$ experiment. {\bf b} Error mitigated decay of increasing weight $Z$ observables, from the simultaneous $T_1$ decay of five qubits (circuit shown in inset). The sign of the odd-weight observables are flipped here, for clarity. Error mitigation of ${\bf c}$ local $\braket{Z_{j-1}Z_j}$ and ${\bf d}$ non-local $\braket{Z_0Z_j}$ observables for 21-qubit GHZ states. Mitigated estimates are obtained by a linear extrapolation of the unmitigated data for stretch factors $c_i=1,1.3,1.6$. The inset of {\bf c} shows the one dimensional chain of 21 qubits used in the experiment, indexed 0 for the top leftmost qubit with increasing indices following the arrow, and the inset of {\bf d} depicts the GHZ state-preparation circuit. The qubit distance for the x-axis of {\bf d} is defined by the difference in qubit index, for non-local observables. All the experiments shown in this figure have ideal expectation values of magnitude +1. The error bars for the error-mitigated results in {\bf{a,b,c,d}} are computed from 50 numerical experiments by bootstrapping the experimental distributions. The individual bootstrapping results are plotted as light red traces for clarity in {\bf{a,b,c,d}}.
}
\label{fig:T1}
\end{figure}

First, we study the performance of zero-noise extrapolation for trivial, multi-qubit product states prepared by simultaneous $T_1$ decay circuits on an initial $\ket{1^{\otimes N}}$ state, see inset of Fig. \ref{fig:T1}(a-b). As shown in the original proposal~\cite{Kristan2017}, the noise can be amplified by stretching the circuit's time evolution, under the assumption of time invariant noise. Here, this amounts to simply stretching the wait times in the $T_1$ decay circuits. In order to reduce the effect of coherence fluctuations ~\cite{kandala2019error,carroll2021dynamics}, all the stretched circuits in this work are averaged together, as discussed previously~\cite{kandala2019error}. Unless otherwise specified, all the experiments discussed in the main text employ 100,000 shots for every stretch factor, and the reported expectation values are corrected for measurement error using a tensor product of individual qubit readout calibration matrices ~\cite{bravyi2020noisy}. Fig.~\ref{fig:T1}(a) depicts the decay of $\braket{Z}$ for the considered stretch factors $c_i = $ 1, 1.5 and 2, and the mitigated observable constructed by linear extrapolation. The mitigated observable shows far superior accuracy over the raw data, even for circuit times going up to 100$\mu$s, comparable to the $T_1$ of the measured qubit. On a line of 5 qubits, we then compare the accuracy of increasing weight $Z$-observables after linear extrapolation, going up to $\braket{Z_0Z_1Z_2Z_3Z_4}$, for varying delay times. As seen in Fig.~\ref{fig:T1}(b), the lower-weight observables display a superior accuracy, for the same state preparation circuit/number of qubits.

We next study the performance of zero-noise extrapolation for low-weight observables for large entangled states. We prepare a GHZ state by applying CNOTs sequentially for 21 qubits along the longest 1D chain in the layout as depicted in Fig.~\ref{fig:T1}(d). There are two sets of observables we examine, the local observables $Z_{j-1}Z_j$ and the non-local observables $Z_0Z_j$ for $j=1,2,\dots,N-1=20$, that all have ideal expectations of $+1$. The CNOT gates are constructed from calibrated cross-resonance (CR) sequences, and all the pulses in the circuit are extended by the considered stretch factors $c=1,1.3,1.6$. 
The circuit structure leads to long idling times, and we see that dynamical decoupling sequences, shown in Fig. \ref{fig:circuit} suppress qubit dephasing and enhance the quality of error mitigation. Fig. \ref{fig:T1}(c-d) illustrates the unmitigated and mitigated expectation values obtained from linear extrapolation. Due to the circuit structure, unmitigated values of these observables suffer $n$-dependent error rates. For example, consider a simple error model where $X$ errors occur after both 2-qubit and 1-qubit gates (including idle locations) in the circuit, but with different probabilities $p_{2}$ and $p_{1}$. We also make the simplifying but largely realistic assumption that qubits in the ground state (i.e.~before any gate is applied) do not suffer errors except for some initialization $X$ error at rate $p_0$. A specific observable is sensitive to errors at only a subset of locations. Generally, if an $X$ error at any of $E_2$ 2-qubit locations, $E_1$ 1-qubit locations, or $E_0$ initializations flips observable $O$, the probability its expectation is $-1$ is 
\begin{align}\label{eq:prob-1}
P_{-1}&=\frac12(1-(1-2p_2)^{E_2}(1-2p_1)^{E_1}(1-2p_0)^{E_0})
\end{align}
and thus $\braket{O}=(1-2p_2)^{E_2}(1-2p_1)^{E_1}(1-2p_0)^{E_0}$. A local observable $Z_{j-1}Z_j$ is sensitive to errors on qubits $j-1$ and $j$ after the CNOT from $j-1$ to $j$ and one of the two initializations, implying $E_0=1$, $E_1+E_2=2(N-j)$, and $E_2=3$ for $j<N-1$ and $E_2=2$ for $j=N-1$. For a non-local observable $Z_0Z_j$, $E_1+E_2=2(N-1)$ is constant with $j$, but $E_2=2+j$ and $E_0=j$. Assuming $p_2>p_1$ and small $N$, we therefore expect a near-linear increase in the local expectations $\braket{Z_{j-1}Z_j}$ with $n$ and an exponential decrease in the non-local expectations $\braket{Z_0Z_j}$. These trends are present in the experimental data
. Consequently, error mitigated observables follow the same trends. This suggests that one should pay careful attention to circuit structure when applying zero-noise extrapolation, and choose to mitigate observables that are less sensitive to noise, such as local observables. 

Finally, we discuss the performance of zero-noise extrapolation on short-depth non-Clifford circuits. As an example, we study the quench dynamics of a transverse-field Ising model. The problem entails studying the evolution of a system under the following Hamiltonian: 
\begin{equation} \label{eq:ising}
    {H}=-J\sum_{\braket{i,j}}{Z_i}{Z_j}+h\sum_i {X_i},
\end{equation}
where $J$ is the strength of the exchange coupling between nearest-neighbor spins with indices $\braket{i,j}$, and $h$ is the transverse magnetic field. The study of quantum quenches with interacting spin-1/2 systems provides a rich playground for explorations of fundamental questions in condensed matter and statistical physics \cite{sachdev1999quantum}, and have already been explored extensively on analog quantum simulators \cite{browaeys2020many,bernien2017probing,zhang2017observation}. These models are also becoming a particularly attractive application of existing, noisy, digital quantum computers \cite{Smith2019,sopena2021,vovrosh2021,urbanek2021} with their increased control and addressability. The spins can be naturally encoded in the physical qubits, nearest-neighbor interactions are accessible with the local-connectivity of the qubits, and relevant quantities can be measured by local, low-weight observables. While this Hamiltonian can be fully diagonalized by a spin to fermion mapping in 1-D and even analytical solutions are known for translationally invariant chains, no such procedure is known for the 2-D model. This makes the 2-D Ising model an attractive testbed for the pursuit of quantum advantage on near-term quantum processors~\cite{childs2018toward}.

\begin{figure*}[hbtp]
\centerline{\includegraphics[width=1.9\columnwidth]{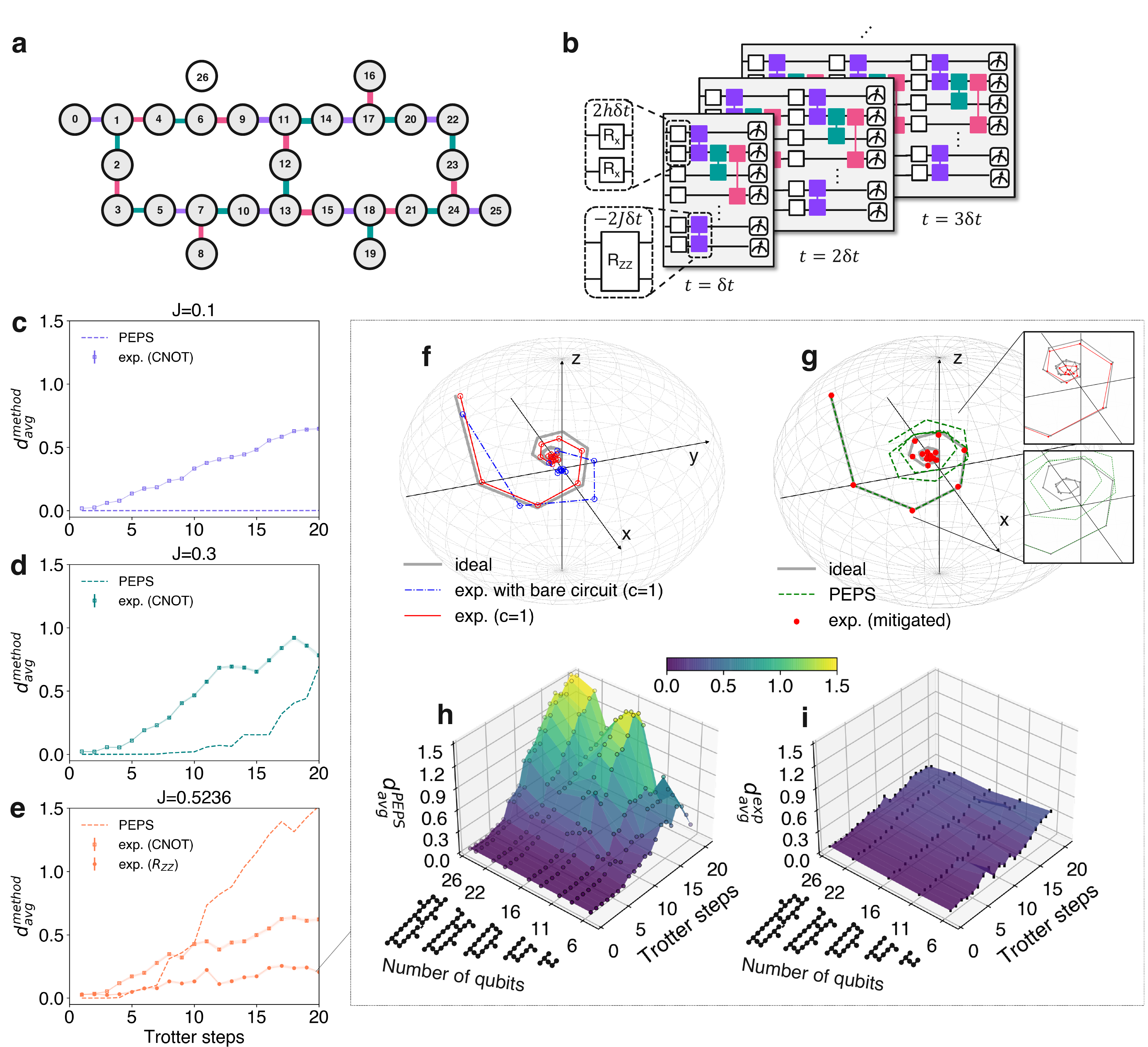}}
\caption{ {\bf{Quench dynamics of 2-D Ising spin lattices: Error mitigated quantum simulation compared to PEPS}} {\bf a} Connectivity of the 26 spin 2-D Ising lattice, mapped onto the physical connectivity of our quantum processor. Color-coding for the links indicates the three sets of simultaneously driven two-qubit gates, based of the device connectivity. {\bf b} Schematic of Trotter circuits for studying the quench dynamics of the Ising Hamiltonian described by Eq.~\ref{eq:Htrotter}. Each Trotter layer is composed of a single qubit gates $R_X(2h\delta t)$ and three sets of simultaneously driven (indicated by color coding) two-qubit gates $R_{ZZ}(-2J\delta t)$. The simulation parameters for the subsequent experiments are transverse field $h=1$, maximum evolution time $T=10$, and total number of Trotter steps $n=20$. At each time step $\delta t$, we measure $\braket{X}_i,~\braket{Y}_i,~\braket{Z}_i$ for the i$^{\textrm{th}}$ qubit, and use it to construct the global magnetization $\vec{M}=(\sum_i\braket{X}_i,\sum_i\braket{Y}_i,\sum_i\braket{Z}_i)/N$. The error in the simulation for each method is quantified by the normalized Euclidean distance  $d_{\textrm{avg}}^{\textrm {method}}=|\vec{M}^{\textrm{ideal}}-\vec{M}^{\textrm{method}}|/|\vec{M}^{\textrm{ideal}}|$, where $\vec{M}^{\textrm {ideal}}$ is obtained from exact numerics and  $\vec{M}^{\textrm {method}}$ is obtained either by the error mitigated experiment, or PEPS. For $N=26$ spins, magnetization error versus number of Trotter steps for varying strengths of exchange coupling, {\bf c} $J=0.1$, {\bf d} $J=0.3$, and {\bf e} $J=0.5236$. Each panel depicts the error from PEPS (dashed line) and the error mitigated experiment with CNOT decomposition of $R_{ZZ}$ gates (squares) from stretch factors $c=1,1.3,1.6$
. For $J=0.5236$, the error from the experiment 
with native gate decomposition of $R_{ZZ}$ (circles) from stretch factors $c=1,1.6,2.0$ is also shown in {\bf e}. {\bf f} Time evolution of the magnetization vector obtained from exact numerics (grey), and unmitigated (c=1) experiment, with (red) and without (blue) the error reduction strategies depicted in Fig.~\ref{fig:circuit} reveals the impact of dynamical decoupling, Pauli twirling and native gate decomposition on the accuracy of the simulation. The origin represents a maximally mixed state. {\bf g} Time evolution of the magnetization vector obtained from exact numerics (grey), error-mitigated experiment with error reduction strategies (red dots), and approximated numeric results using PEPS algorithm for bond dimension $D=4$ (green). The mitigation involves linear extrapolation of observables measured at stretch factors $c=1,1.6,2.0$. Insets highlight the superior accuracy of the error mitigated trajectory, in comparison to PEPS. Magnetization error versus number of Trotter steps and smaller 2-D spin lattices, for {\bf h} PEPS and {\bf i} the error mitigated experiment. The spread in the error-mitigated results in {\bf{c,d,e,i}} from finite sampling, is estimated using 50 numerical experiments obtained by bootstrapping of the experimentally measured distributions. For clarity, individual bootstrapping results are plotted as lighter traces in {\bf{c,d,e}}, and are within the marker size in {\bf i}. }
\label{fig:quench}
\end{figure*}

Here, we study the quench dynamics of a 2-D spin lattice that follows the connectivity of our processor, excluding a single inferior qubit. Starting in the ground state of the non-interacting Hamiltonian ($J=0$), with the qubits initialized in $\ket{\psi_0}=\ket{0^{\otimes 26}}$, the time evolution of the system is studied once the exchange coupling is turned on suddenly at time $t=0$.
\begin{equation} \label{eq:ising}
    \ket{\psi(t)}=e^{-i{H}t}\ket{\psi_0}.
\end{equation} 

We implement the dynamics on our hardware by a first-order Trotter decomposition of the the time evolution.
\begin{equation} \label{eq:Htrotter}
    \begin{split}
    e^{-i{H}t} =&  e^{-i({H}_Z+{H}_X) t}  \\
    \approx & \prod_{1}^n e^{-i{H}_Z \delta t}e^{-i{H}_X \delta t}, \\
    \end{split}
\end{equation}
where $H_Z=-J\sum_{\braket{i,j}}{Z_i}{Z_j}$ and $H_X=h\sum_i {X_i}$. The maximum evolution time $T$ is discretized into time steps $\delta t= T/n$, with $n$ Trotter layers, and
\begin{equation} \label{eq:HzHx}
    \begin{split}
        e^{-i{H}_Z \delta t}=&\prod_{\braket{i,j}}\exp\left\{i(J\delta t) {Z_i}{Z_j}\right\} \\
        e^{-i{H}_X \delta t}=&\prod_{i}\exp\left\{-i(h\delta t) {X_i}\right\}. \\
    \end{split}
\end{equation}
Unless otherwise specified, the experiments described here fix $h=1$, $T=10$, $\delta t=0.5$ and study the dynamics for varying $J$. The evolution can now be implemented via a combination of single-qubit rotations and $ZZ$ two-qubit gates between pairs of connected qubits on the lattice. We study the dynamics for up to 26 spins on a 2-D lattice and $n=20$ Trotter steps, where Fig.~\ref{fig:quench}(a-b) illustrates each Trotter layer implements 3 blocks of parallelized two-qubit gates on next-nearest neighbors, for a maximum circuit depth of 60. 

Although a comparison to the ideal result, obtained by {\it{exact}} numerical simulation, is possible for sizes up to 26 spins, a mere increase in system size by a factor of 2-3 renders this unfeasible. At this larger scale, only approximate classical methods are available. It is important to point out here, that the quantum experiment itself only provides an approximation to the idealized simulation due to the hardware noise. To establish the validity of approximate quantum computation with error mitigation, it is therefore prudent to compare to established approximate classical methods that scale efficiently. 

Simulating time dynamics of general quantum systems is a class of decision problems solvable in polynomial time by a quantum computer, $BQP$-complete~\cite{nielsen_chuang_2010}, and no classical approximation method is therefore expected to produce a universally accurate approximation. However, a wide range of approximate classical simulation methods \cite{moyal1949quantum,keldysh1965diagram,white2004real} have been developed that provide good approximations in various limiting cases. Starting from a limiting case where the approximation works well, we want to provide a comparison with the quantum approximation and see for which parameter values the experimental expectation values provide a higher accuracy. Here, we focus on a comparison with tensor network methods since our experiment considers a locally interacting spin Hamiltonian. In particular, since we consider the dynamics of 2-D systems, we compare the experimental results to the standard projected entangled pair state (PEPS) method \cite{verstraete2004renormalization,murg2007variational,verstraete2008matrix}. These states are the current prime candidate to capture ground state properties of gapped 2-D spin systems. This method can also be used to approximate dynamics of weakly entangled systems, but it is generally expected that the PEPS simulation will become inaccurate when strong entanglement is produced \cite{schuch2008entropy}. 

Here, we consider the original PEPS time evolution algorithm \cite{murg2007variational} with bond dimensions $D=4$. The algorithm scales as ${\cal O}(D^{12} N)$ in time and ${\cal O}(D^8)$ in memory and can become numerically unstable for larger bond dimensions. At each Trotter step, we estimate the average magnetization of the interacting spin system by a measurement of the weight-1 observables averaged over 26 qubits - $\braket{X},\braket{Y},\braket{Z}$. We quantify and compare to the experimental error in terms of the normalized Euclidean distance between the magnetization vectors obtained from PEPS and the ideal vector. Fig.~\ref{fig:quench}(c-e) show that PEPS exhibits excellent agreement with the exact numerics over the entire range of the Trotterized evolution when the $R_{ZZ}(\theta)$ gates are weakly entangling for $J=0.1$, but displays increasingly poorer accuracy with increasing $J$ and trotter steps. 

With our digital approach, the time evolution for different coupling parameters can be experimentally studied by simply varying the angle of the single qubit rotations $R_X(2h\delta t)$ and the two qubit gates $R_{ZZ}(-2J\delta t)$. Arbitrary  $R_{ZZ}(\theta)$ gates can be implemented on our device by using 2 CNOT gates and additional single qubit rotations, where the CNOT gates are themselves constructed from cross-resonance driven $R_{ZX}(\pi/2)$ interactions that are native to our hardware. This enables us to compare the performance of zero-noise extrapolation on circuits with the same structure and evolution time, while systematically increasing the amount of entanglement with increasing $J$. Error mitigation is ultimately limited by the noise in the circuit, and in order to extend its reach, we adopt additional error reduction strategies such as dynamical decoupling, previously discussed for GHZ circuits. Furthermore, an important source of crosstalk in fixed-frequency architectures is the $ZZ$ interaction~\cite{wei2021quantum}, that can be an important limitation to circuit fidelity, and particularly problematic for circuits simulating Ising Hamiltonians. In this context, Pauli twirling~\cite{knill2004fault,wallman2016} is an attractive approach to average out off -diagonal coherent errors of the circuit in the Pauli basis and improve circuit fidelity~\cite{hashim2021randomized}. It also enables the suppression of coherent errors in the gate-rescaling ~\cite{kandala2019error} that could otherwise lead to unphysical extrapolations. Here, each two-qubit gate is sandwiched between twirling gates and inverting gates such that the same unitary is implemented, and for two-qubit Clifford operations such as CNOT gates, the twirling gates can be sampled randomly from a set of tensor products of all single-qubit Pauli gates. By incorporating these error suppression strategies with linear extrapolation, we estimate the magnetization, and the corresponding error for $J=0.1,~0.3,~0.5236$ in Fig.~\ref{fig:quench}(c-e). 
Interestingly, the accuracy of the error mitigated magnetization does not reveal a strong dependence on the degree of entanglement in experiment. As a consequence, our error mitigated experiment provides an increasingly competitive approximation to the magnetization over PEPS, for increasing $J$. Therefore, we focus on the case of $J=0.5236$, that specifically requires the implementation of $R_{ZZ}(\pi/6)$ gates for our chosen Hamiltonian and Trotterization parameters. 

With CNOT decomposition, our longest circuit times exceed 80 $\mu$s, limiting the accuracy of the mitigated observables. For $J=0.5236$, the performance of our experiment can be further enhanced by employing a more time-effective gate decomposition that uses shorter cross-resonance pulses to calibrate partially entangling~\cite{kandala2019error,earnest2021pulseefficient} $R_{ZX}(\pi/6)$ gates. Details of the calibration and benchmarking are discussed in the supplementary section. With this decomposition, we achieve a $\sim4\times$ reduction in circuit time. These considerations and strategies highlight the benefit of pulse-level stretching for error amplification and the associated freedom over choice of stretch factors. For large volume circuits, error amplification via the insertion of identity equivalent gates can lead to circuit times beyond the coherence budget.  Additionally, we show that a Pauli twirling strategy can also be employed with the non-Clifford, fractional $R_{ZZ}(\theta)$ gates, albeit with a limited set of twirling gates, see Supplementary Information. Remarkably, even for these large volume circuits, we observe improvements in circuit performance for a very modest number of twirled circuit instances $\sim$ 8, see Supplementary Information. 

The impact of the combined error suppression strategies is visualized in Fig.~\ref{fig:quench}(f) by tracking the evolution of the magnetization vector. Furthermore, the error mitigated trajectory  displays excellent qualitative agreement with the ideal noise-less path in Fig.~\ref{fig:quench}(g), while the trajectory obtained from PEPS diverges at increasing Trotter steps.  With the optimal gate decomposition, we see that the error mitigated experiment provides a closer approximation to the magnetization than PEPS even for Trotter steps exceeding $n=6$ for the largest value of $J=0.5236$ considered in this work, Fig.~\ref{fig:quench}(e). However, we also note that the accuracy of larger-weight observables, after error mitigation, is not as competitive, see Supplementary information. We also compare the magnetization error for increasing qubit number (and varying 2D lattices), and map this for increasing numbers of Trotter steps in Fig. \ref{fig:quench}(h),(i). We see that our experiment is closer to the ideal simulation than PEPS for a large part of this phase space, particularly for increasing qubit number and Trotter steps.  These observations of accurate coherent dynamics in error mitigated expectation values going up to 26 qubit circuits with depth $d=60$, in conjunction with the scaling of the PEPS ($D=4$) error shown in Fig. \ref{fig:quench}(i), raises the enticing possibility of studying quench dynamics on noisy, digital, fully-programmable quantum processors that are beyond exact diagonalization as well as efficient, classical approximate techniques.

The experiments discussed here highlight several important considerations to access competitive computations from noisy, intermediate-sized quantum processors. We demonstrate the scalability of zero noise extrapolation and show that a number of experimental strategies can be employed to further enhance its performance on a noisy device. Efforts to improve the noise scaling can enable higher order Richardson that provide error mitigated expectation values with an even higher degree of accuracy, that we expect to be continually supplemented by improvements in the coherence and quality of quantum hardware. Already, with existing error rates and linear extrapolation, we see that error mitigated magnetization dynamics of 2-D Ising quenches for 26 spins can provide a better approximation than PEPS at ($D=4$) in the limit of increasing entanglement. In this context, the development of 65 qubit processors based on our hardware platform~\cite{zhang2020}, with comparable error rates and 2-D connectivity, makes these devices an attractive platform for the exploration of error mitigated quantum dynamics of 2-D spin systems that are classically intractable.

\bibliographystyle{naturemag}


{\bf Acknowledgments}
We thank Isaac Lauer for contributions towards two-qubit gate calibration, and Douglas T. McClure, Neereja Sundaresan for valuable discussions on device bring-up. 

{\bf Author information} 
The authors declare no competing financial interests. Correspondence and requests for materials should be addressed to Y.K. (youngseok.kim1@ibm.com) or A. K. (akandala@us.ibm.com).

\widetext

\begin{center}
\newpage
\textbf{\large Supplementary Information: Scalable error mitigation for noisy quantum circuits produces competitive expectation values}
\end{center}

\setcounter{equation}{0}
\setcounter{figure}{0}
\setcounter{table}{0}
\setcounter{page}{1}
\setcounter{section}{0}
\makeatletter
\renewcommand{\theequation}{S\arabic{equation}}
\renewcommand{\thefigure}{S\arabic{figure}}
\renewcommand{\thesection}{S\Roman{section}}
\renewcommand{\citenumfont}[1]{S#1}

\section{Device characteristics}\label{sec:device}
Our processor is a heavy-hex network of 27 single-Josephson junction transmon qubits with fixed qubit-qubit couplings with all-microwave control. Single qubit gates are implemented using microwave pulses with Gaussian envelope and DRAG corrections~\cite{motzoi2009simple}. Two qubit entangling gates are also implemented with microwave pulses, employing the cross-resonance interaction~\cite{paraoanu:2006, chow2011}. Table~\ref{tab:kolkata} summarizes the single-qubit metrics for the devices, with daily measurements over 15 days. 
\begin{table}[h!]
\centering

 \begin{tabular}{||c c c c c||} 
 \hline
  & median & mean & min & max \\ [0.5ex] 
 \hline\hline
 $f_{01}$ (GHz) & 5.10 & 5.07 $\pm$ 0.11 & 4.87 & 5.26 \\ 
 $f_{01}-f_{12}$ (MHz) & 343.41 & 344.82 $\pm$ 7.31 & 338.86 & 378.87 \\ 
 $T_1$ ($\mu s$) & 116.50 & 115.29 $\pm$ 25.92 & 54.48 & 168.16 \\ 
 $T_2$ ($\mu s$) & 106.85 & 106.33 $\pm$ 59.65 & 20.67 & 215.40 \\
 1Q EPC ($\times 10^{-4}$) & 2.23 & 2.92 $\pm$ 2.93 & 1.40 & 16.02 \\
 readout infidelity (\%) & 0.93 & 1.06 $\pm$ 0.60 & 0.51 & 3.09 \\ [1ex]
 \hline
 \end{tabular}
 \label{tab:kolkata}
\caption{{\bf{Summary of single qubit properties on $ibmq\_kolkata$}}. Reported $T_1$, $T_2$, 1Q EPC, readout infidelity were obtained from daily measurements over a 15 day period.}

\end{table}

\begin{figure}[h]
\centerline{\includegraphics[width=0.8\columnwidth]{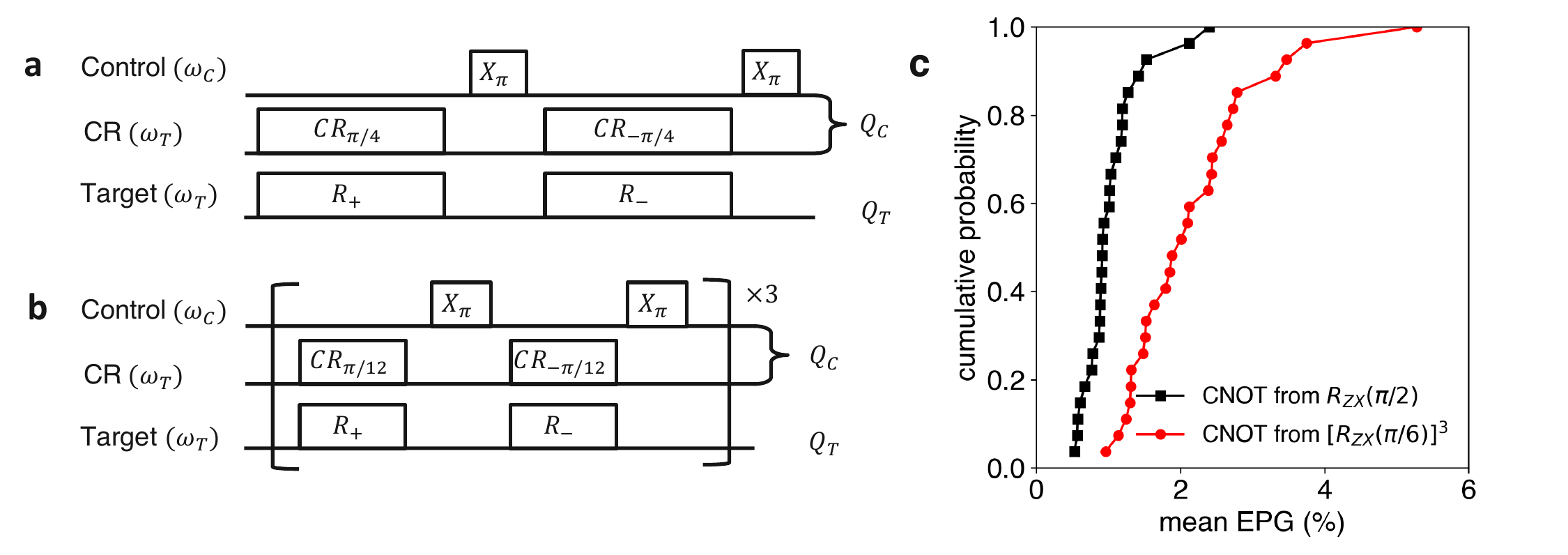}}
\caption{{\bf{Two-qubit gate calibration and benchmarking}} Pulse sequences for calibration of {\bf a} $R_{ZX}(\pi/2)$ and {\bf b} $R_{ZX}(\pi/6)$ gates. {\bf c} Summary of CNOT gate fidelities across the device for constructions from $R_{ZX}(\pi/2)$ sequences (black) or $[R_{ZX}(\pi/6)]^3$ sequences (red).}
\label{fig:partial_gate}
\end{figure}

Two-qubit CNOT gates are constructed from cross-resonance (CR) pulses and single qubit rotations. The $ZX$ interaction is isolated by using an echoed-CR sequence on the control qubit $Q_C$ with rotary drives on the target qubit $Q_T$, and the gates are typically calibrated to a $R_{ZX}(\pi/2)$ for CNOT construction. The pulse sequence was previously reported in ~\cite{Sundaresan2020} and is depicted in Figure~\ref{fig:partial_gate}(a). The fractional $R_{ZZ}(\theta)$ gates employed for the quench dynamics were constructed from $R_{ZX} (\theta)$ sequences. Specifically the $R_{ZX}(\theta=\pi/6)$ gates considered here were calibrated to a $R_{ZX}(\pi/2)$ by repeating the gate three times, as illustrated in Figure~\ref{fig:partial_gate}(b). The width of the CR pulses for the $R_{ZX}(\theta=\pi/6)$ were chosen to have calibrated amplitudes that were similar or smaller than the amplitudes employed for CNOT calibration, with unchanged rise-fall profiles. With the fractional gates, the rise-fall sections form a larger fraction of the pulse width, and the single qubit pulses form a larger fraction of the gate time. Hence, the gate time does not exactly scale with the fraction of the original $\pi/2$ conditional rotation. All the two-qubit gates are characterized by randomized benchmarking of CNOT gates constructed from either $R_{ZX}(\pi/2)$ sequences or $[R_{ZX}(\pi/6)]^3$ sequences, and a summary of device-wide gate fidelities are shown in Figure~\ref{fig:partial_gate}(c).

\section{Experimental considerations for optimal performance of error mitigation} \label{subsec:DD_PT}
The performance of quantum error mitigation strategies are ultimately limited by the noise in the circuit of interest. In addition to reducing the circuit time by optimal gate decomposition, detailed in Fig.~\ref{fig:circuit} and Sec.~\ref{sec:device}, we elaborate on important experimental considerations to further suppress errors and enhance our signal-to-noise. The importance of these considerations is illustrated by their impact on the quench circuits discussed in the main text.

\subsection{Dynamical decoupling}
\begin{figure}[!b]
\centerline{\includegraphics[width=1\columnwidth]{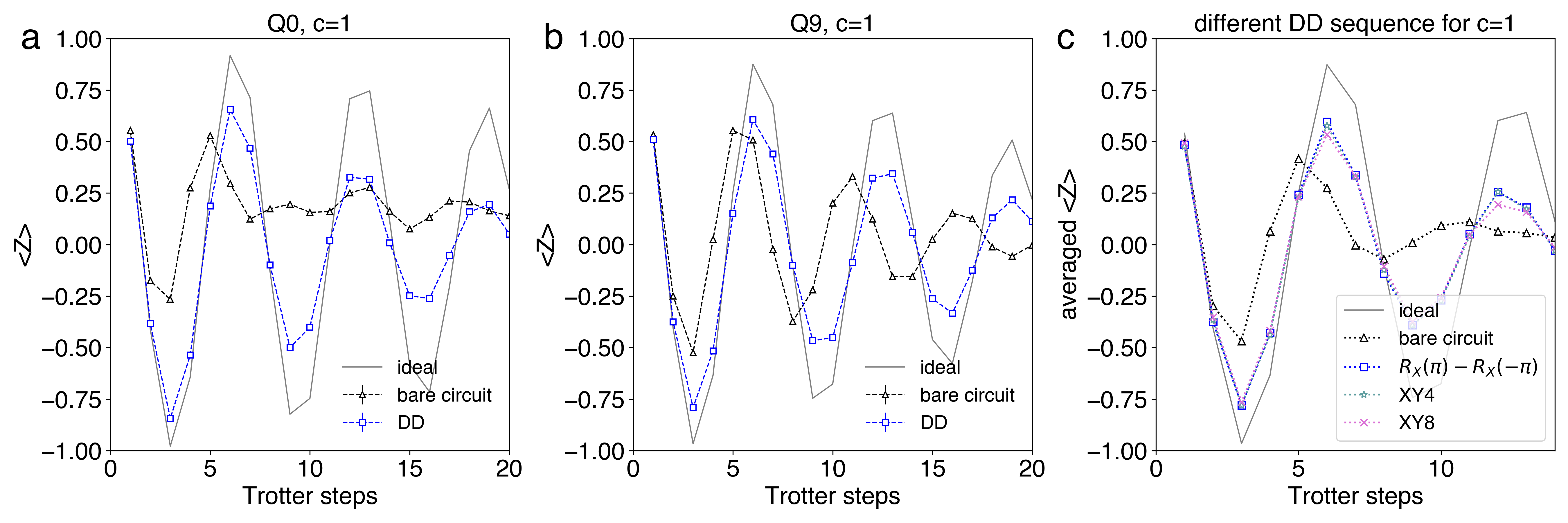}}
\caption{{\bf{Impact of Dynamical decoupling on the performance of quench circuits}} The panels depict data from the quench dynamics with 26 qubits, for $J=0.1$. Unmitigated $\braket{Z}$ observables, measured with and without DD, are compared to the ideal evolution for individual qubits {\bf{a}} Q0 and {\bf{b}} Q9, and reveal the impact of DD in extending coherent evolution to longer circuit depths, and refocusing coherent error, respectively.  {\bf{c}} Comparison of various DD sequences with the ideal evolution of the average $\braket{Z}$. The sequence examined here includes $\tau/4-R_{X}(\pi)-\tau/2-R_{X}(-\pi)-\tau/4$ (denoted as $R_{X}(\pi)-R_{X}(-\pi)$), $\tau/8-R_{X}(\pi)-\tau/4-R_{Y}(\pi)-\tau/4-R_{X}(\pi)-\tau/4-R_{Y}(\pi)-\tau/8$ (denoted as XY4)\cite{Maudsley1986}, and $\tau/16-R_{X}(\pi)-\tau/8-R_{Y}(\pi)-\tau/8-R_{X}(\pi)-\tau/8-R_{Y}(\pi)-\tau/8-R_{Y}(\pi)-\tau/8-R_{X}(\pi)-\tau/8-R_{Y}(\pi)-\tau/8-R_{X}(\pi)-\tau/16$ (denoted as XY8)\cite{Gullion1990}, where $\tau$ is the idling time of a qubit minus length of the total single qubit pulses in a given DD sequence. When the qubit idling time is shorter than the total DD sequence length, we use shorter DD sequence (e.g. XY4 instead of XY8). No discernible improvement is observed by increasing the number of decoupling pulses beyond the $R_{X}(\pi)-R_{X}(-\pi)$ sequence, which we then employ through the rest of this work.}
\label{fig:quench_DD}
\end{figure}

For the GHZ and quench circuits considered in this work, the physical connectivity and circuit structure leads to idling times for qubits, that can be a source of dephasing and coherent errors from undesired classical and static $ZZ$ crosstalk. We see that some of these errors can be refocused by dynamical decoupling (DD) sequences~\cite{Viola1999,ZANARDI1999} that have been previously employed on our hardware to improve circuit fidelity~\cite{Jurcevic_2021}. We insert the following sequence $\tau/4-R_{X}(\pi)-\tau/2-R_{X}(-\pi)-\tau/4$ for every idling period of time, where $\tau$ is idling time minus the length of two $R_X(\pi)$ pulses. We highlight the impact of DD on the quench dynamics for $J=0.1$ with the $c=1$ circuits in Fig.~\ref{fig:quench_DD}. The figure shows specific examples of DD extending the circuit depth out to which clear coherent evolution of a qubit can be observed (Fig.~\ref{fig:quench_DD}(a)), as well as the suppression of coherent error (Fig.~\ref{fig:quench_DD}(b)). This demonstrates the importance of DD to improve the ``bare" signal and enhance the performance of zero-noise extrapolation. We also compared the performance of this sequence with other standard variants in Fig.~\ref{fig:quench_DD}(c), and did not observe discernible improvements for our considered circuits. 

\subsection{Pauli twirling}
\begin{figure}[!t]
\centerline{\includegraphics[width=1\columnwidth]{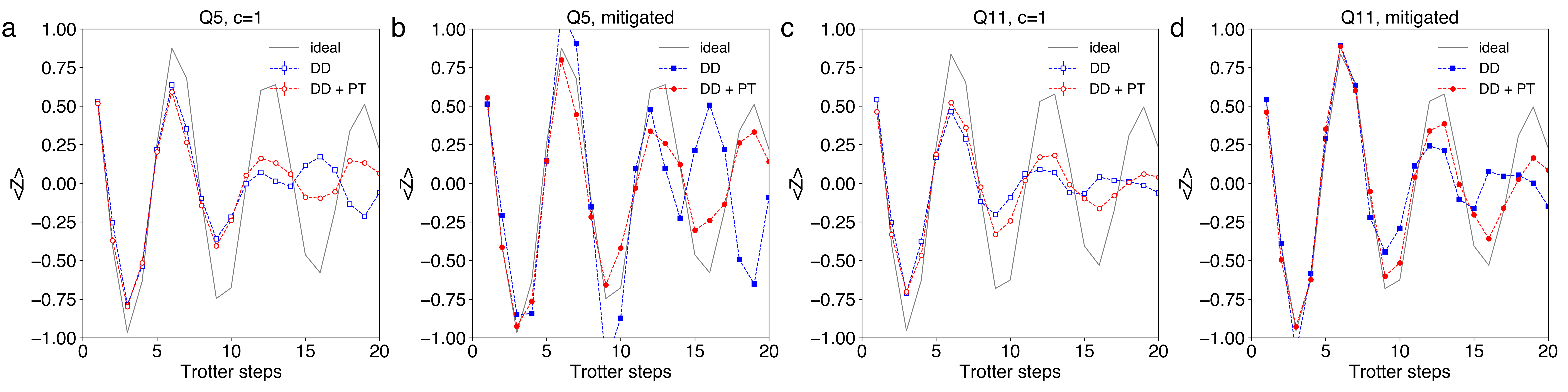}}
\caption{{\bf{Impact of Pauli Twirling on the performance of quench circuits}}
The panels depict data from the quench dynamics with 26 qubits, for $J=0.1$. Linear extrapolation is performed using stretched gates with $c=1,1.3,1.6$, and DD sequences are inserted for idling qubits. All experiments employed a total of 100,000 shots, per stretch factor. The experiments with Pauli Twirling (PT) use 4 random twirling instances, with 25,000 shots for each instance. The experimentally measured dynamics of the $\braket{Z}$ magnetization is compared to the ideal evolution, with and without Pauli twirling (PT), for individual qubits without mitigation in {\bf{a}} Q5 and {\bf{c}} Q11, and with mitigation in {\bf{b}} Q5 and {\bf{d}} Q11.}
\label{fig:quench_PT}
\end{figure}

Although the dynamical decoupling (DD) sequences can help with refocusing some coherent errors, we find Pauli twirling~\cite{knill2004fault} to be another important ingredient for further enhancing circuit performance. Fig.~\ref{fig:quench_PT}(a) shows examples of discernible coherent error in the results of post-DD quench circuits ($J=0.1$). These coherent errors can also be source of inaccurate noise scaling, leading to even unphysical expectation values after error mitigation, as seen in Fig.~\ref{fig:quench_PT}(b). Pauli twirling is a powerful technique to convert an arbitrary noise channel into a stochastic Pauli error channel, suppressing off-diagonal coherent error contributions. This is implemented by sandwiching Clifford gates between randomly sampled twirling gates such that the net operation, in the absence of noise, is unchanged. The twirling gates inserted before the Clifford operation are typically sampled from a complete set of Pauli operations, $\mathbb{G}=\{I,X,Y,Z\}^{\otimes N}$ for $N$ qubits, and an appropriate set of Pauli operations are then applied after the Clifford gate for a net equivalent unitary operation. For our quench circuits with the CNOT decomposition of $R_{ZZ}(\theta)$ operations, each CNOT (Clifford) gate is sandwiched between gates randomly sampled from $\mathbb{G}$ and Pauli gates that preserve the net CNOT operation, and the expectation values are then averaged over multiple twirling instances. For each CNOT, the twirling gates are combined with other single qubit operations in the circuit, to avoid any increase in circuit time. These circuit randomizations can be particularly helpful for suppressing the amplification of coherent error in structured circuits with repetitive gate layers such as the Trotter circuits considered in this work. Fig.~\ref{fig:quench_PT}(a-b) shows that the incorrect oscillation period as well as inaccurate noise scaling are corrected using twirling gates. Fig.~\ref{fig:quench_PT}(c-d) show another example that after twirling, the oscillations of the measured observable are ``in-phase" with the ideal evolution for the raw data and extrapolation results.

\begin{figure}[!h]
\centerline{\includegraphics[width=1\columnwidth]{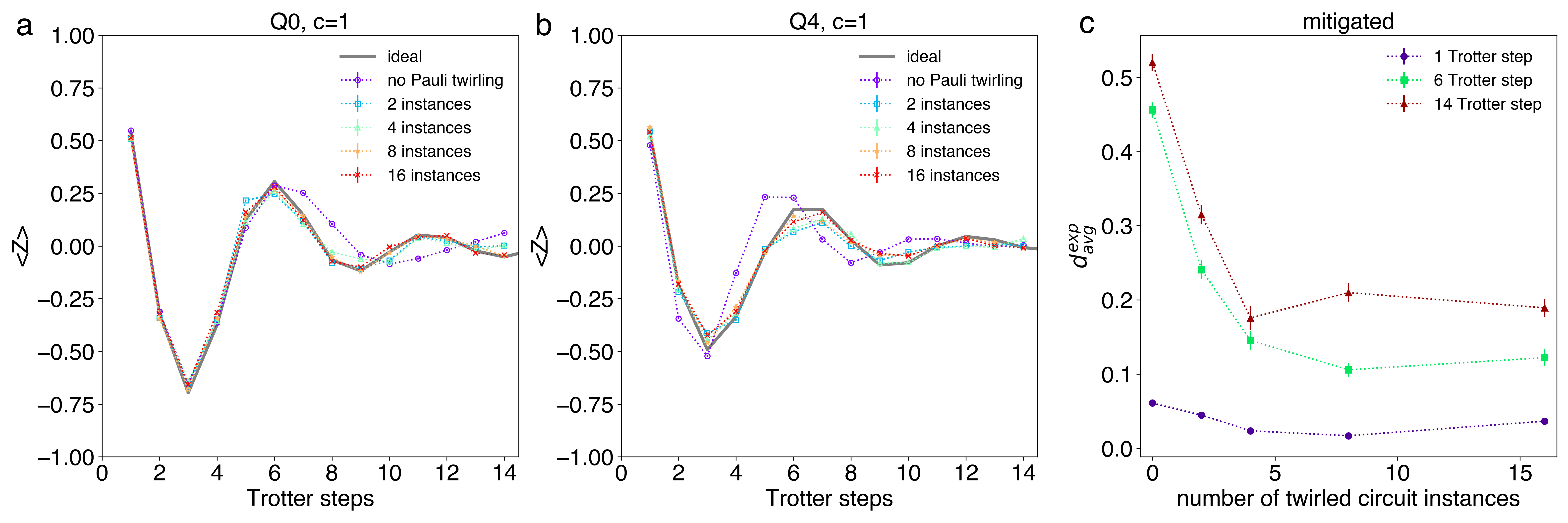}}

\caption{{\bf{Impact of Pauli twirling for quench circuits with native gate decomposition of $R_{ZZ}(\theta)$ gates. }} The panels depict data from the quench dynamics with 26 qubits, for $J=0.5236$, with native gate decomposition of $R_{ZZ}(\theta)$ gates. Linear extrapolation is performed using stretched gates with $c=1,1.6$, and DD sequences are inserted for idling qubits. All experiments employed a total of 60,000 shots, per stretch factor. The experimentally measured dynamics of the $\braket{Z}$ magnetization is compared to the ideal evolution, for a number of Pauli twirling (PT) instances, for individual qubits without mitigation in {\bf{a}} Q0 and {\bf{c}} Q4. {\bf{c}} For three different trotter steps, the normalized Euclidean distance  of the error mitigated magnetization vector from the ideal vector $d_{avg}^{exp}$ shows no further improvement beyond 8 twirling instances. The error bars for the error-mitigated results in {\bf{c}} are computed over 50 numerical experiments by bootstrapping the experiment data with the identical number of shots.}
\label{fig:quench_PT2}
\end{figure}

The native gate decomposition for $R_{ZZ}(\theta)$ significantly shortens the circuit time but involves non-Clifford gates, and here as well, the bare quench circuits ($J=0.5236$) are plagued by coherent error as shown in Fig.~\ref{fig:quench_PT2}(a,b). For this decomposition, we randomly sample twirling gates on the [control,target] pair from a subset $\mathbb{G}_{ZZ} \subseteq \mathbb{G}$ that commute with $R_{ZZ}(\theta)$, $\mathbb{G}_{ZZ}=\{ [I,I],[X,X],[Y,Y],[Z,Z] \}$. For every draw, the same set of twirling gates are applied before and after the $R_{ZZ}(\theta)$, in order to implement the same unitary. Even with this very limited twirling set, we observe very significant improvements in the performance of our circuits, shown in Fig.~\ref{fig:quench_PT2}(a,b). Indeed, it was shown \cite{cai2019constructing} that for specific noise models it is sufficient to only twirl over a smaller subset of Pauli matrices to average out the off-diagonal error.

A final important consideration here has to do with the number of twirling instances. Our longest circuits at depth 60 employ 540 $R_{ZZ}$ gates and with twirling gates sampled from $\mathbb{G}_{ZZ}$ for each $R_{ZZ}$, a natural question has to do with the number of instances required to effectively suppress our coherent errors. Remarkably in Fig.~\ref{fig:quench_PT2}(c), for the native decomposition and the quench dynamics with $J=0.5236$, we see no further improvement in the Euclidean distance from the ideal magnetization beyond a mere 8 twirled circuit instances.

\subsection{Noise model dependence of the identity insertion noise amplification method}
Following \cite{dumitrescu2018} many error mitigation experiments \cite{He2020,lowe2020,sopena2021,urbanek2021} have used the idea of inserting identity gates in to the circuit by repeating gates that square to the identity an odd number of times. The most frequently considered example is the insertion of an odd number, e.g. $2k+1$, of $\mbox{CNOT}$ gates. This approach to noise amplification can be seen as an alternative to the stretching of gate times. The intent is to achieve an amplification of the noise parameter by a factor $2k+1$.  A particular example where this can work is when the noisy gate is simply given by a perfect $\mbox{CNOT}$ gate preceded by a uniform $2$ qubit depolarizing channel with depolarizing error $\epsilon$. If we write for the error $(1-\epsilon) = \exp(-\lambda)$ we see that the repeated application of the noisy gate acts on the depolarizing error as $(1-\epsilon)^{2k+1}$ which translates to the desired amplification of the noise rate to $(2k+1) \lambda$. The obvious appeal of this approach lies in the fact that this method is considerably easier to implement in hardware than the stretching of gates which need to be precisely calibrated. The gate insertion can be directly implemented on the circuit level without any changes to the underlying experiment.\\ 

However, we caution that the method can provide noise amplification results that arbitrarily deviate from the desired result when only slightly more complicated noise models are considered. To provide a simple example, consider the direct generalization of the $2$- qubit depolarizing channel to a Pauli - error noise model. The Pauli channel $\Lambda$ acts on a state $\rho$ by applying some Pauli matrix $P_{\alpha} \in \langle 1,X_i,Y_i,Z_i \rangle_{i=1\ldots n}$, where $\alpha$ denotes a $2n$ bit-string, with probability $q_\alpha$.  The Pauli channel is then given by
\[
\Lambda(\rho) = \sum_{\alpha = 0}^{4^n-1} q_\alpha P_\alpha \rho P_\alpha^\dagger.
\]
To closely resemble the depolarizing case above, we assume that this noise model is applied right before the $\mbox{CNOT}$ gate leading to a noise controlled-not gate described by the channel ${\cal CNOT}(\rho) = \mbox{CNOT}~\Lambda(\rho)~\mbox{CNOT}^\dagger$. A convenient property of the Pauli noise model that can be verified by direct computation, is that it acts on Pauli matrices by multiplication. That is the Pauli matrices form an eigenbasis of the channel with eigenvalues $\mu_\alpha$. Hence, we have that the channel acts as $\Lambda(P_{\alpha}) = \Lambda^*(P_{\alpha}) = \mu_\alpha P_\alpha$ in both the Schroedinger $\Lambda$ and Heisenberg $\Lambda^*$ picture. The desired noise amplification procedure that is consistent with the extrapolation has to map for $2k+1$ applications the eigenvalues $\mu_\alpha$ to $\mu_\alpha^{2k+1}$. This does not happen for Pauli channels as we will see here. Since the $\mbox{CNOT}$ gate is a Clifford gate, it maps Pauli matrices to other Pauli matrices so that $\mbox{CNOT}~P_\alpha~\mbox{CNOT}^\dagger =  P_{\alpha_{cx}}$ where $\alpha$ and $\alpha_{cx}$ are related by a linear transformation determined by the $\mbox{CNOT}$ gate. Let us now look at the expectation value of a Pauli matrix $P_\alpha$ we want to mitigate.  In the example here assume the expectation value is obtained by applying a single $\mbox{CNOT}$ to an initial state $\rho_0$. We obtain
\[
	\tr \left[ P_{\alpha}{\cal CNOT}(\rho_0) \right] = \mu_{{\alpha_{cx}}} \tr \left[P_{\alpha_{cx}} \rho_0 \right]
\]
then following the identity insertion trick the noise is amplified to 
\[
	\tr \left [ P_{\alpha} {\cal CNOT}^3(\rho_0) \right ] = \mu_{{\alpha_{cx}}}^2\mu_{\alpha} \tr \left[ P_{\alpha_{cx}} \rho_0 \right].
\]
The correct noise amplification for zero noise extrapolation however asks that the noise is amplified by $\mu_{{\alpha_{cx}}} \rightarrow \mu_{{\alpha_{cx}}}^3$, which is different from $\mu_{{\alpha_{cx}}}^2 \mu_{\alpha}$. For Pauli noise channels with non-uniform values of $\mu_{\alpha}$ strong deviations from the desired behavior can be constructed. This example shows that even for relatively simple and only slightly more realistic noise channels the identity insertion method can provide an amplification of the noise that can lead to uncontrolled estimates when applying the zero noise extrapolation method.  

\subsection{Two level systems} \label{sec:TLS}

\begin{figure}[!t]
\centerline{\includegraphics[width=1\columnwidth]{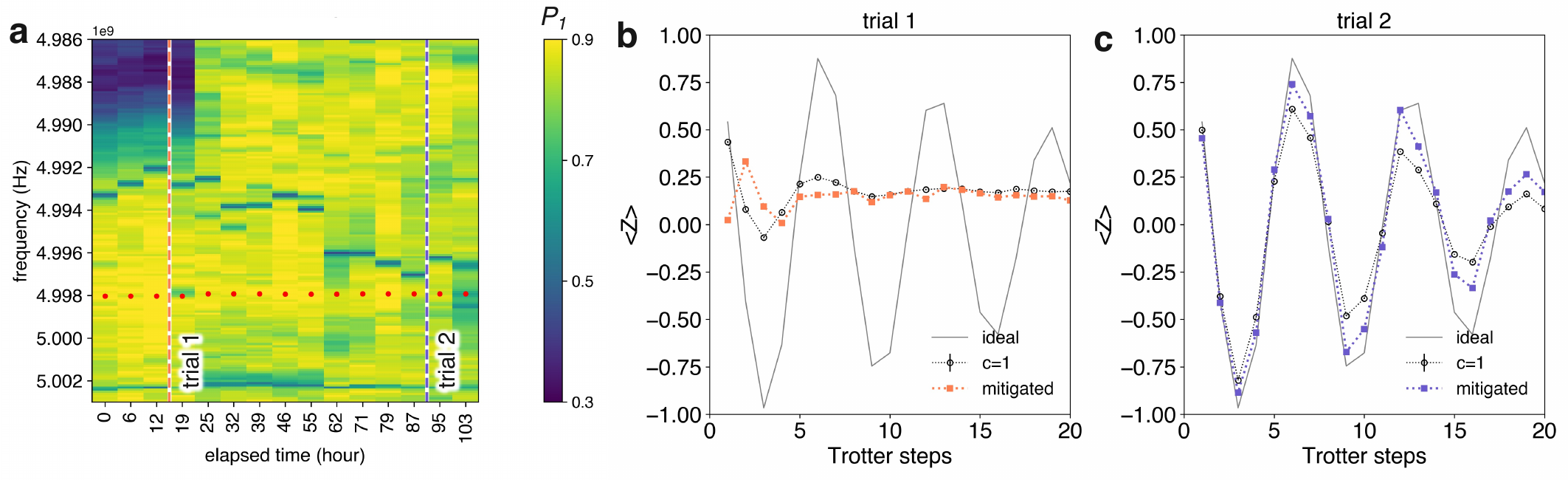}}
\caption{{\bf{Impact of Two level system (TLS) fluctuations on the performance of quench circuits}} {\bf{a}} TLS spectroscopy of  Q23. The spectroscopy tracks the probability of excited state population ($P_1$) at a fixed $T_1$ delay time of $20\mu s$ over a range of Stark shifted frequencies, after an excited state is prepared~\cite{carroll2021dynamics}. The bare qubit frequency is indicated as a red dotted line, and data is collected every 6-7 hours to monitor temporal and spectral fluctuations in $P_1$.  The quantum quench experiment is carried out at two separate times, indicated by the vertical dotted lines. The impact of the TLS environment on the error mitigated quench dynamics for 26 spins, is shown for for $J=0.1$. Linear extrapolation is performed using stretched gates with $c=1,1.3,1.6$, and DD sequences are inserted for idling qubits. All experiments employed a total of 100,000 shots, per stretch factor, with 4 twirling instances. {\bf{b}} The TLS at trial 1 leads to a suppressed coherent evolution of the qubit's $\braket{Z}$ magnetization, seen by comparison to the ideal evolution. {\bf{b}} At trial 2, the qubit recovers its coherent evolution of $\braket{Z}$, for both the unmitigated c=1, as well as the mitigated data.}
\label{fig:quench_TLS}
\end{figure}

An important challenge for the stability of our experiments in general, and the performance of zero-noise extrapolation more specifically, is associated with resonant interactions of our qubits with defect two-level systems (TLSs), that lead to fluctuations in coherence ~\cite{Klimov2018,carroll2021dynamics} and system performance. As recently shown for single-junction transmons~\cite{carroll2021dynamics}, we map the TLS spectral environment and spectral dynamics in the vicinity of the qubit frequency by measuring the qubit energy relaxation time $T_1$ at ac-Stark shifted frequencies every 6-7 hours. In order to speed up the spectroscopy, we use the excited state probability $P_1$ at a fixed delay time as a proxy for $T_1$. We focus on the TLS dynamics of a single qubit Q23 in Fig.~\ref{fig:quench_TLS}(a) and correlate it with the performance of quench circuits for $J=0.1$. Blue regions in the Fig.~\ref{fig:quench_TLS}(a) represent spectral dips in $P_1 (T_1)$ that correspond to TLS resonances. We focus on circuit performance at two time instances, referred to trial 1 and trial 2. For trial 1, we see a strong TLS in the spectral vicinity of the operating qubit frequency, and Fig.~\ref{fig:quench_TLS}(b) shows the strongly suppressed coherent oscillations in the quench dynamics, seen both in the $c=1$ data as well as the mitigated result. The strongly inhomogeneous noise profile in the vicinity of the qubit frequency further complicates the performance of zero noise extrapolation with CR gates. At the different stretch factors, the driven control qubit is Stark shifted in frequency by different amounts, and this can potentially violate assumptions on time invariant noise. For instance, the control qubit Stark shift during CR23-24 (control-target) operation for $c=1 (=1.6)$ is $+9.71 (+2.87)$ MHz. However, at instance trial 2, the absence of a strongly coupled TLS leads to far improved performance in the raw, as well as the error mitigated data, Fig.~\ref{fig:quench_TLS}(c).

\section{local weight-2 observables in quench dynamics of 2D Ising spin lattices} \label{sec:zz_observables}
\begin{figure}[!h]
\centerline{\includegraphics[width=0.8\columnwidth]{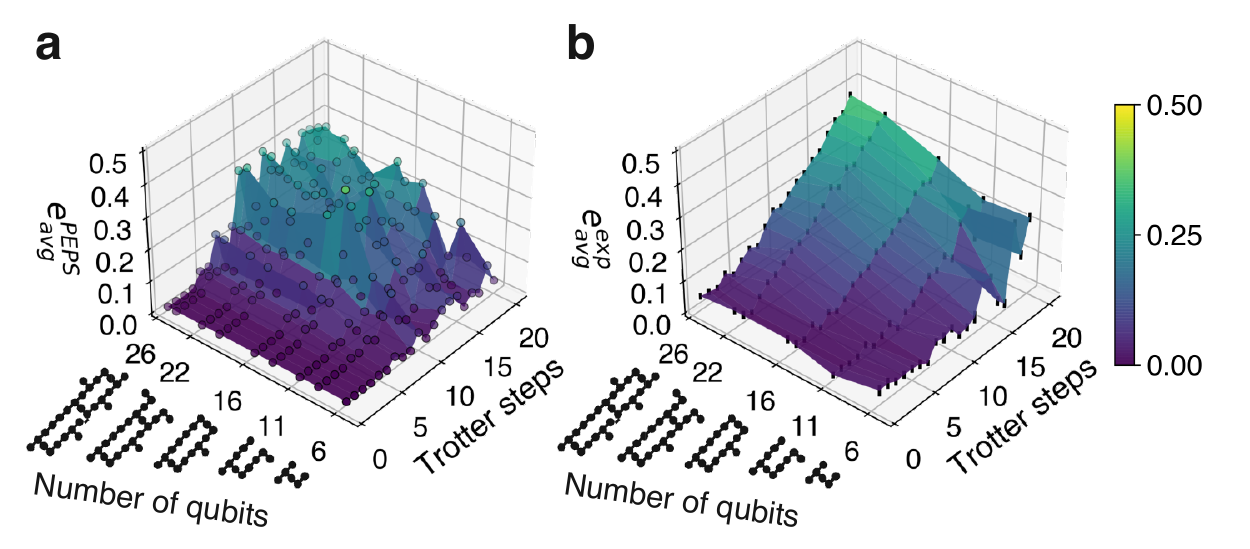}}
\caption{{\bf{Quench dynamics of 2-D Ising spin lattices ($J=0.5236$): Comparison of experimental and PEPS error for $\braket{ZZ}$}} {\bf a} Relative error from PEPS, $e_{avg}^{PEPS}$ defined in Sec.~\ref{sec:zz_observables} versus number of Trotter steps and size of the 2-D spin lattice. The PEPS simulation is performed with the bond dimension $D=4$. {\bf b} Relative error from the error mitigated experiment, $e_{avg}^{exp}$ defined in Sec.~\ref{sec:zz_observables} versus number of Trotter steps and size of the 2-D spin lattice. The data is taken from the same experiment described in Fig.~\ref{fig:quench}(i).}
\label{fig:quench_weight2}
\end{figure}

As discussed in the main text for the $T_1$ and GHZ circuits, large weight and non-local observables can be increasingly sensitive to noise, which in turn affects the efficacy of zero-noise extrapolation. While our results in Fig.~\ref{fig:quench} of the main text focus on weight-1 observables for quench dynamics, this naturally raises the question of experimental performance for higher weight observables as well. Here, we focus on $\braket{ZZ}$ as an example of local, weight-2 observables. We obtain the averaged local ZZ observable as $\braket{ZZ}=\sum_{\braket{i,j}} \braket{Z_iZ_j}/N_{adj}$, where $\braket{i,j}$ represents the adjacent qubit pairs and $N_{adj}=27$ is the total number of adjacent pairs for 26 qubits in the layout described in Fig.~\ref{fig:quench}(a). We measure an error of the obtained $\braket{ZZ}$ from \emph{exact} numerical results by defining a relative error $e_{avg}^{method}=|\braket{ZZ}^{ideal}-\braket{ZZ}^{method}|/|\braket{ZZ}^{ideal}|$, where $\braket{ZZ}^{method}$ is obtained either from PEPS or error mitigated experiment. Fig.~\ref{fig:quench_weight2} compares the relative PEPS error with the error mitigated experiment for 
$J=0.5236$. 
While the error mitigated experiment provides better approximations for weight-1 observables as illustrated in Fig.~\ref{fig:quench}(h-i) over PEPS ($D=4$), we do not observe such a definitive win for weight-2, local $\braket{ZZ}$ expectation values that are increasingly sensitive to noise in the circuit than their weight-1 counterparts.

\section{Individual qubit results for quench dynamics of a 2-D Ising spin lattice} \label{sec:ind_bits}


\begin{figure*}[hbtp]
\centerline{\includegraphics[width=1\columnwidth]{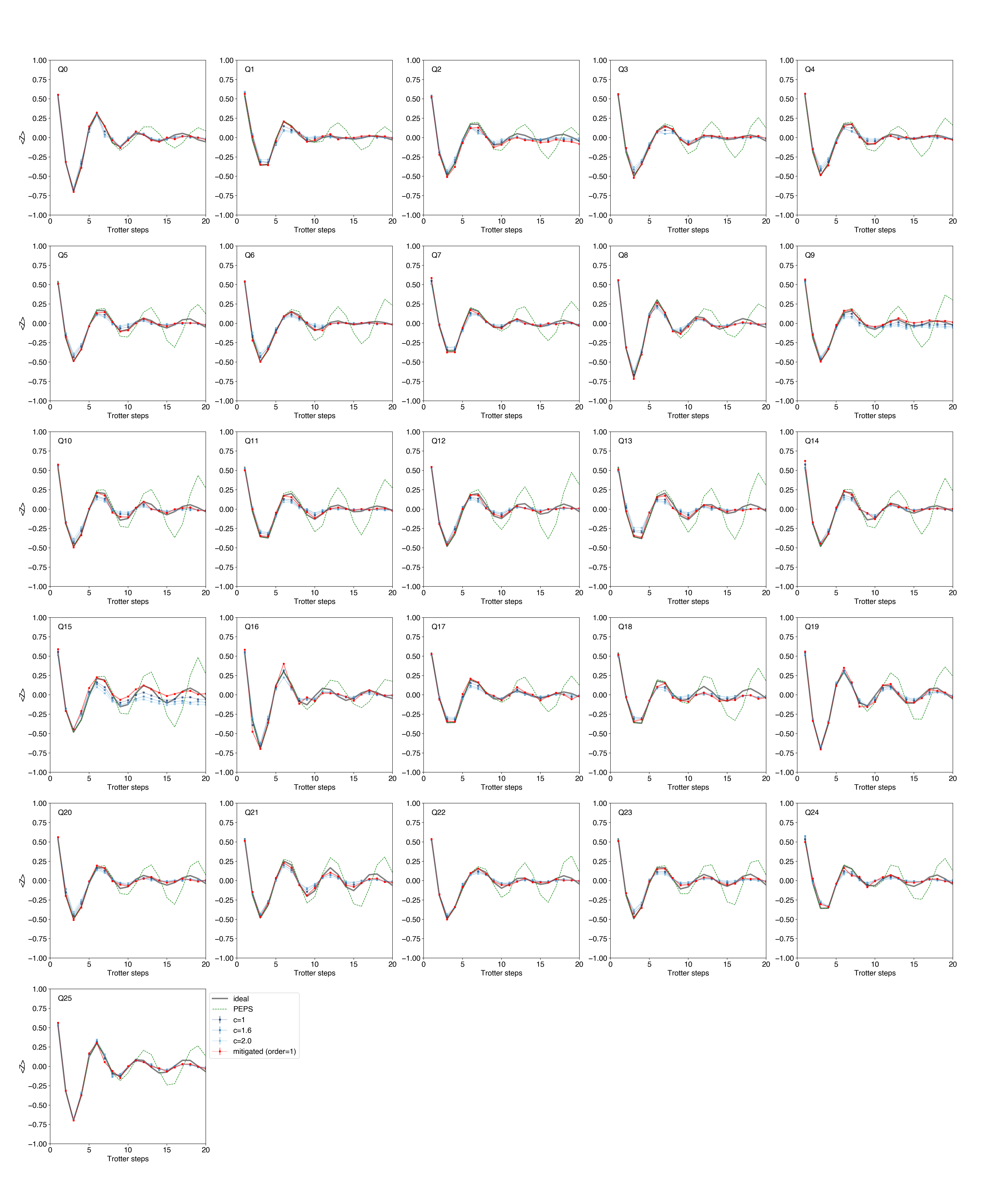}}
\caption{{\bf{Individual qubit $\braket{Z}$ evolution for quench dynamics with $J=0.5236$.}} The qubit index follows labels indicated in Fig.~\ref{fig:quench}(a). The expectation value is measured from 100,000 shots at every stretch factor. A DD sequence is inserted for idling qubits and we use 8 random twirling gate instances. The error-mitigation is performed from linear extrapolation using stretch factors $c=1,1.6,2$. We also plot exact numerical results (grey solid line) and numerical approximation using PEPS method with bond dimension $D=4$ (green dashed line) for comparison.}
\end{figure*}

\begin{figure*}[hbtp]
\centerline{\includegraphics[width=1\columnwidth]{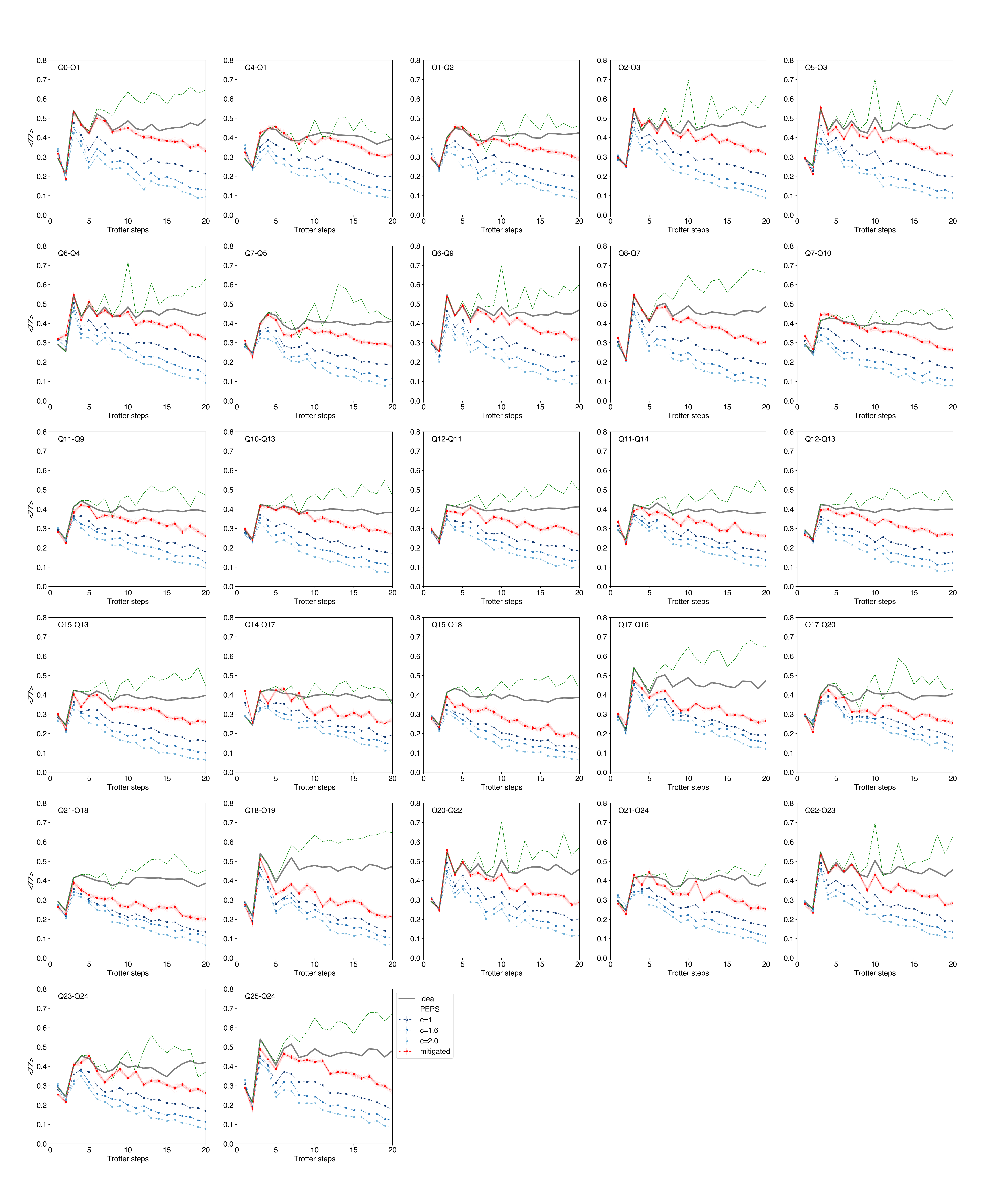}}
\caption{{\bf{Quench experiment for individual local $\braket{ZZ}$ observable at $J=0.5236$.}} The qubit index follows labels indicated in Fig.~\ref{fig:quench}(a). The local $\braket{ZZ}$ is computed for adjacent qubit pairs. The observable is computed from 100,000 shots. A DD sequence is inserted for idling qubits and we use 8 random twirling gate instances. The error-mitigation is performed from linear extrapolation using stretch factors $c=1,1.6,2$. We also plot exact numerical results (grey solid line) and numerical approximation using PEPS method with bond dimension $D=4$ (green dashed line) for comparison.}
\end{figure*}

\end{document}